# Competition between charge-density-wave and superconductivity in the kagome metal RbV$_3$Sb$_5$

N. N. Wang[1,2=], K. Y. Chen[1,2=], Q. W. Yin[3=], Y. N. N. Ma[1,2], B. Y. Pan[4,1], X. Yang[1,2], X. Y. Ji[1,2], S. L. Wu[1,2], P. F. Shan[1,2], S. X. Xu[1,2], Z. J. Tu[3], C. S. Gong[3], G. T. Liu[1,2], G. Li[1,2], Y. Uwatoko[5], X. L. Dong[1,2], H. C. Lei[3*], J. P. Sun[1,2*], and J.-G. Cheng[1,2*]

[1]Beijing National Laboratory for Condensed Matter Physics and Institute of Physics, Chinese Academy of Sciences, Beijing 100190, China

[2] School of Physical Sciences, University of Chinese Academy of Sciences, Beijing 100190, China

[3] Department of Physics and Beijing Key Laboratory of Opto-electronic Functional Materials & Micro-nano Devices, Renmin University of China, 100872 Beijing, China

[4] School of Physics and Optoelectronic Engineering, Ludong University, Yantai, Shandong 264025, China

[5] Institute for Solid State Physics, University of Tokyo, Kashiwa, Chiba 277-8581, Japan

= These authors contributed equally to this work.

*E-mail: jpsun@iphy.ac.cn, hlei@ruc.edu.cn, jgcheng@iphy.ac.cn

## Abstract

The interplay between charge-density-wave (CDW) order and superconductivity (SC) in the Kagome metal RbV$_3$Sb$_5$ is studied by tracking the evolutions of their transition temperatures, $T^*$ and $T_c$, as a function of pressure ($P$) via measurements of resistivity and magnetic susceptibility under various hydrostatic pressures up to ~ 5 GPa. It is found that the CDW order at $T^*$ experiences a subtle modification at $P_{c1} \approx 1.5$ GPa before it is completely suppressed around $P_{c2} \approx 2.4$ GPa. Accordingly, the superconducting transition $T_c(P)$ exhibits a shallow M-shaped double superconducting dome with two extrema of $T_c^{onset} \approx 4.4$ K and 3.9 K around $P_{c1}$ and $P_{c2}$, respectively, leading to a fourfold enhancement of $T_c$ with respect to that at ambient pressure. The constructed $T$-$P$ phase diagram of RbV$_3$Sb$_5$ resembles that of CsV$_3$Sb$_5$, and shares similar features as many other unconventional superconducting systems with intertwined competing electronic orders. The strong competition between CDW and SC is also evidenced by the broad superconducting transition width in the coexistent region. Our results shed more light on the intriguing physics involving intertwined electronic orders in this novel topological kagome metal family.

## Introduction

In the last decades, numerous research efforts have been devoted to exploring the interplay between superconductivity (SC) and various competing electronic orders in unconventional superconducting systems [1-10]. The recently discovered Kagome metals $A$V$_3$Sb$_5$ ($A$ = K, Rb, and Cs) fall into this category because they show the coexistence of charge-density-wave (CDW) order and SC in addition to the presence of nontrivial topological band structure [11-14]. The superconducting transition occurs at $T_c$ = 0.93 K, 0.92 K and 2.5 K, while the CDW order appears at $T^* \approx$ 78 K, 104 K and 94 K for KV$_3$Sb$_5$ [13], RbV$_3$Sb$_5$ [14] and CsV$_3$Sb$_5$ [12], respectively, as revealed by resistivity, magnetization, specific heat, x-ray diffraction, optical spectroscopy and scanning tunneling microscopy measurements [12-24]. Interestingly, a chiral CDW state with breaking time-reversal symmetry has been observed in $A$V$_3$Sb$_5$ and was considered to be critical for the emergence of giant anomalous Hall effect and unconventional SC [15,25]. Moreover, theoretical calculations on Kagome Hubbard model with different electron filling states have shown many exotic phases, including spinless fermions [26], valence-bond solid phases [27], CDW state [28,29], chiral spin-density-wave (SDW) state [28,30], exotic superconducting states [28-32] and topological point defects [33]. In order to elucidate the detailed interplay between the intertwined CDW order and SC, it is important to tune the electronic states of $A$V$_3$Sb$_5$ by methods such as chemical doping, intercalation, or applying high pressures.

So far, several high-pressure studies have been performed on KV$_3$Sb$_5$ and CsV$_3$Sb$_5$ to unveil the intimated correlations between the CDW order and SC [34-39]. For KV$_3$Sb$_5$, the application of high pressure was found to enhance the superconducting transition temperature $T_c$ in concomitant with the suppression of the CDW order, suggesting a strong competition between CDW and SC [36]. However, for CsV$_3$Sb$_5$ with much larger interlayer distance, detailed high-pressure transport measurements reveal a more complex relationship between CDW and SC, displaying an unusual M-shaped double superconducting dome accompanying a monotonic suppression of CDW order [34,39]. Such an unusual phase diagram of CsV$_3$Sb$_5$ should arise from a subtle modification of the CDW order associated with the large compression of interlayer distance as indicated by the DFT calculations [34]. The more dispersive band structure along the $c$ axis for KV$_3$Sb$_5$ with much reduced interlayer distance (or $c$-axis) [11,40] can explain the rapid suppression of CDW order at a lower critical pressure of $P_c \approx$ 0.4 GPa [36]. Moreover, resistance measurements on CsV$_3$Sb$_5$ by using diamond anvil cells over a much extended pressure range have uncovered the emergence of a second superconducting phase (SC-II) above 15 GPa with a maximum $T_c \approx$ 5 K [35,37,38,41]. Since the high-pressure XRD rules out the occurrence of structural phase transition around this pressure [37], the observed SC-II phase at $P$ > 15 GPa has been attributed to a Lifshitz transition as supported by the transport measurements and band structure calculations [35,37]. In comparison with K and Cs, Rb has an intermediate atomic radius and consequently the interlayer $c$-axis distance of $RbV_3Sb_5$ lies between that of $KV_3Sb_5$ and $CsV_3Sb_5$ [14]. It is thus interesting to investigate the evolutions of CDW and SC in $RbV_3Sb_5$ under hydrostatic pressures in order to gain a comprehensive understanding on the relationship between CDW order and SC in this class of Kagome superconductors.

In this work, we have performed detailed resistivity, direct-current field (dc) and alternate-current field (ac) magnetic susceptibility measurements on $RbV_3Sb_5$ single crystal by using the piston-cylinder cell (PCC) and cubic anvil cell (CAC) under various hydrostatic pressures up to 5.2 GPa. Our results reveal a shallow M-shaped double superconducting dome in $RbV_3Sb_5$, which should correlate with the cryptic modification of the CDW order at $P_{c1} \approx 1.5$ GPa before it is completely suppressed around $P_{c2} \approx 2.4$ GPa. Moreover, the maximum $T_c$ can be enhanced to ~ 4.4 K at about 1.5 GPa, a fourfold enhancement compared with $T_c$ at ambient pressure. The constructed $T$-$P$ phase diagram, similar with that of $CsV_3Sb_5$ [34], clearly reveals the competition between CDW and SC, providing more insights into the high-pressure properties of this topological kagome superconducting family.

**Experimental details**

Single crystals of $RbV_3Sb_5$ were synthesized by Rb ingot (purity 99.9%), V powder (purity 99.9%) and Sb grains (purity 99.999%) using the self-flux method [14]. Temperature dependences of resistivity and ac magnetic susceptibility for $RbV_3Sb_5$ samples were measured simultaneously by using a self-clamped PCC under various hydrostatic pressures up to 2.2 GPa [42]. Here, we use the Daphne 7373 as the pressure transmitting medium (PTM) in PCC. The resistivity was measured with standard four-probe method with the electrical current applied within the $ab$-plane. The magnetic field was applied along the $c$-axis. The ac susceptibility of $RbV_3Sb_5$ together with a piece of Pb placed in the same coil was measured with the mutual induction method. An excitation current of ~ 1 mA with a frequency of 1117 Hz was applied to the primary coil and the output signal was picked up with a Standford Research SR830 lock-in amplifier. The measured superconducting transition of Pb was used to determine the pressure value in PCC and it was also used as a reference to estimate the superconducting shielding volume of the $RbV_3Sb_5$. We have also employed a palm-type CAC to measure the resistivity of $RbV_3Sb_5$ up to 5.2 GPa [43]. Glycerol was employed as the liquid PTM for CAC. Finally, we used a miniature BeCu PCC to measure the dc magnetization under various pressures up to 0.84 GPa in the commercial magnetic property measurement system (MPMS-3, Quantum Design). The $RbV_3Sb_5$ crystals together with a piece of lead (Pb) were loaded into a Teflon capsule filled with Daphne 7373 as the PTM, and the pressure value was determined from the relative shift of the $T_c$ of Pb. All the measurements were performed in the zero-field-cooled mode.

**Results and Discussions**

Figure 1(a) shows the temperature dependence of in-plane resistivity $\rho(T)$ of RbV$_3$Sb$_5$ single crystal at ambient pressure. The normal-state $\rho(T)$ exhibits a typical metallic behavior with the residual resistivity ratio $RRR = \rho(290\ K)/\rho(1.5\ K) = 39$, indicating a high quality of our samples. As can be seen, a kink-like anomaly appears in $\rho(T)$ at $T^* \approx 103$ K as indicated by the downward arrow, and this feature is correlated to the formation of the CDW order [14]. An enlarged view of $\rho(T)$ below 1.5 K is depicted in the inset of Fig. 1(e), which shows that the superconducting transition starts at around 1.1 K and reaches zero resistance at about 0.78 K. Here, the $T_c^{onset}$ is determined as the interception between two straight lines below and above the superconducting transition and $T_c^{zreo}$ is defined as the zero-resistivity temperature. These results are consistent with the previous report [14]. Then, we measure the field dependence of resistivity $\rho(H)$ of RbV$_3$Sb$_5$ at various temperatures up to 0.93 K with the field applied along the $ab$ plane and the $c$-axis, respectively, as shown in Figs. 1(b) and 1(c). We can see that the superconducting upper critical field $\mu_0H_{c2}$ is continuously shifted to lower fields with increasing temperature gradually. Here, we determined the upper critical field, $\mu_0H_{c2}$, from the 90% drops of $\rho(H)$ curves and plotted the temperature dependence of $\mu_0H_{c2}(T)$ in Fig. 1(d). As can be seen, the $\mu_0H_{c2}(T)$ can be well fitted by using the Ginzburg-Landau (G-L) formula: $\mu_0H_{c2}(T) = \mu_0H_{c2}(0)(1-t^2)/(1+t^2)$, where $\mu_0H_{c2}(0)$ is defined as zero-temperature upper critical field and $t$ represents the reduced temperature $T/T_c$. The calculated $\mu_0H_{c2}^{//ab}(0)$ and $\mu_0H_{c2}^{//c}(0)$ are 0.3 T and 0.11 T, respectively. Moreover, the corresponding G-L coherent lengths are estimated to be $\xi^{ab}_{GL} = 547.0$ Å and $\xi^{c}_{GL} = 200.6$ Å based on the formula: $\mu_0H_{c2}^{//c}(0) = \Phi_0/2\pi\xi^{ab}_{GL}{}^2$ and $\mu_0H_{c2}^{//ab}(0) = \Phi_0/2\pi\xi^{ab}_{GL}\xi^{c}_{GL}$, where $\Phi_0 = hc/2e$ is the magnetic flux quantum [44]. Therefore, the obtained anisotropy parameter is $\gamma = \mu_0H_{c2}^{//ab}(0)/\mu_0H_{c2}^{//c}(0) \approx 2.73$, which is about one third of that in CsV$_3$Sb$_5$ [45]. The reduced anisotropy is consistent with the reduced ionic radius of alkali metal from Cs to Rb. For the anisotropic superconductors, we can use the ratio $\gamma = \sqrt{m_c^*/m_{ab}^*}$ to express the band-structure anisotropy, where the $m_c^*$ and $m_{ab}^*$ are the effective mass of the quasiparticles along the $c$ axis and within the $ab$ plane, respectively. The estimated $m_c^*/m_{ab}^* \sim 7.5$ indicates a relatively strong anisotropy of the band structure in RbV$_3$Sb$_5$.

To further characterize the superconducting transition of RbV$_3$Sb$_5$, we measured the low-temperature magnetization $M(T)$ at a magnetic field of 5 Oe under zero-field-cooled (ZFC) and field-cooled (FC) conditions. As shown in Fig. 1(e), the obvious diamagnetic signal can be seen in the ZFC and FC curves, and it reveals the bulk SC after correcting the demagnetization factor. The onset of superconducting transition appears at $T_c \approx 0.78$ K, consistent with the $T_c^{zreo}$ determined from the $\rho(T)$ data shown in the inset of Fig. 1(e). Figures 1(f) and 1(g) present the field dependences of magnetization $M(H)$ up to 50 Oe at various temperatures from 0.39 to 0.6 K along the $ab$-plane and $c$-axis, respectively. Apparently, the large magnetic hysteresis characterizes the common behavior of type-II superconductor. A linear fitting to $M(H)$ for the full shielding effect yields the lower critical field $\mu_0H_{c1}$. Here, the obtained lower critical field values are $\mu_0H^{//ab}_{c1}(0) = 9.42$ Oe and $\mu_0H^{//c}_{c1}(0) = 4.7$ Oe by employing the G-L formula as displayed in Fig. 1(h). Furthermore, according to the equations $H_{c1}^{//c}(0) = (\Phi_0/4\pi\lambda_{ab}^2)\ln(\kappa_c)$, and $H_{c1}^{//ab}(0) = (\Phi_0/4\pi\lambda_{ab}\lambda_c)\ln(\kappa_{ab})$, where the G-L parameters $\kappa_c = \lambda_{ab}/\xi_{ab}$ and $\kappa_{ab} = \sqrt{\lambda_{ab}\lambda_c/\xi_{ab}\xi_c}$ [46], we can further estimate the penetration depth to be $\lambda^{ab}_{GL} = 551.7$ Å and $\lambda^c_{GL} = 201.4$ Å. The calculated $\gamma = \lambda_{ab}/\lambda_c \sim 2.73$ is consistent with the value estimated above, and the G-L parameters $\kappa^{ab}_{GL} = \lambda^{ab}_{GL}/\xi^{ab}_{GL} = 1.01$ and $\kappa^c_{GL} = \lambda^c_{GL}/\xi^c_{GL} = 1$, larger than $1/\sqrt{2}$, further confirm that RbV$_3$Sb$_5$ belongs to the type-II superconductor.

Figures 2(a) and 2(b) display the temperature dependences of resistivity $\rho(T)$ and its derivative $d\rho/dT$ below 120 K under various pressures up to 2.2 GPa measured with a PCC. Here, we shift the $\rho(T)$ and $d\rho/dT$ curves vertically for clarity. The evolution of the CDW ordering temperature with pressure can be tracked from the resistivity anomaly. At 0 GPa, the $\rho(T)$ shows a kink-like anomaly at $T^* \approx 103$ K (Fig. 1(a) and Fig. 2(a)). From $d\rho/dT$, we can actually define two characteristic temperatures, i.e. $T^{peak}$ and $T^{dip}$ corresponding to the peak and dip temperatures, and determine the $T^* = (T^{peak} + T^{dip})/2$. With increasing pressure gradually, the anomaly in $\rho(T)$ at $T^*$ and the corresponding $T^{peak}$ and $T^{dip}$ in $d\rho/dT$ continuously move to lower temperatures. Interestingly, the peak feature is diminished gradually with pressure, and the $T^{peak}$ and $T^{dip}$ are merged to about 53 K at 1.5 GPa, above which the dip feature in $d\rho/dT$ becomes less obvious. The weakening of the anomaly in resistivity has a profound influence on the superconducting transition as we will discuss in detail below. It is noteworthy that the kink anomaly in $\rho(T)$ at $T^*$ changes to a hump-like feature with increasing pressure (Fig. 2(a)). Similar feature has also been observed in CsV$_3$Sb$_5$ under pressure [34,39]. Due to the quasi-2D nature of the $A$V$_3$Sb$_5$ family, it is inevitable that the interlayer interactions will be enhanced upon reducing the ionic radius of $A$ cations. As shown in our previous work, the CDW order involves a non-vanishing order wave-vector along the $c$-axis [34], which can explain the hump-like feature in the $\rho_c(T)$ of RbV$_3$Sb$_5$ at ambient pressure [14].

An enlarged view of the low-temperature $\rho(T)$ data under various pressures are present in Fig. 3(a). At ambient pressure, the superconducting transition cannot be detected down to $T = 1.5$ K, the lowest temperature in our high-pressure measurements. When increasing pressure gradually, we start to see the weak drop of $\rho(T)$ at 0.43 GPa, and then an obvious superconducting transition at $T_c^{onset} \approx 2.5$ K at 0.76 GPa. But zero-resistivity cannot be achieved down to 1.5 K. $T_c^{onset}$ and $T_c^{zero}$ rapidly rise to ~ 3.5 K and ~ 1.6 K at 1.02 GPa, and then further increase to ~ 4 K and ~ 2.7 K at 1.29 GPa, respectively. In this pressure range, the superconducting transition is broad with a transition width $\Delta T_c$ over 1.5 K; such a broad transition is consistent with the fact that SC coexists with the CDW in this regime. With increasing pressure to 1.5 GPa, the superconducting transition temperature reaches a maximum with $T_c^{onset} \approx 4.4$ K and $T_c^{zero} \approx 4$ K; accordingly, the superconducting transition width quickly shrinks to $\Delta T_c \approx 0.4$ K. Interestingly, when the pressure is further increased from 1.5 to 2.19 GPa, $T_c^{onset}$ and $T_c^{zero}$ are reduced slightly to ~ 4.1 K and ~ 3.55 K, respectively. In the pressure range 1.76 -2.19 GPa, $\Delta T_c$ increases slightly to ~ 0.6 K. The broadened $\Delta T_c$ highlights a complex and intrinsic phenomenon that may originate from the cryptic modification of the CDW as discussed below.

To further track the evolution of $T_c(P)$ under higher pressures, we measured the $\rho(T)$ of RbV$_3$Sb$_5$ up to 5.2 GPa with CAC, and display the low-temperature data in Fig. 3(b). The $\rho(T)$ in the whole temperature range are given in Fig. S1. The $\rho(T)$ at 1.9 GPa in CAC resembles those of 1.76 GPa and 2 GPa in PCC, showing a relatively sharp superconducting transition with $T_c^{onset} \approx 4.14$ K and $T_c^{zero} \approx 3.66$ K. When the pressure is increased to 2.4 GPa, the normal-state resistivity is reduced significantly, and the $T_c^{zero}$ ($T_c^{onset}$) increases (decrease) slightly to ~3.8 (3.93) K, resulting in a very sharp transition with $\Delta T_c \approx 0.1$ K. Above 2.8 GPa, the superconducting transition shifts to lower temperatures monotonically, and the $T_c^{onset}$ and $T_c^{zero}$ are reduced to 2.06 K and 1.96 K at 5.2 GPa. Here, we can see that the superconducting transition remains very sharp with $\Delta T_c \approx 0.1$ K at $P \geq 2.4$ GPa. These results indicate that the observed broadening of the superconducting transition in the intermediate pressure range is not due to the sample or pressure inhomogeneity, but is an intrinsic property. These $\rho(T)$ measurements have thus revealed a complex, non-monotonic variation of $T_c(P)$ in the investigated pressure range.

The evolution of the superconducting transition was further monitored by measuring the dc magnetization $M(T)$ up to 0.84 GPa in MPMS and the ac susceptibility $\chi'(T)$ at various pressures to 2.2 GPa in PCC. Figure 3(c) shows the zero-field-cooled $M(T)$ data measured under an external magnetic field of $H = 5$ Oe in the warming-up process. The diamagnetic signal in $M(T)$ appears at $T_c^\chi \approx 2$ K for 0.84 GPa, where the resistivity data shows a remarkable superconducting transition. Figure 3(d) shows the $\chi'(T)$ data at various pressures up to 2.19 GPa and the results are in good agreement with the resistivity data shown in Fig. 3(a). Upon applying pressure gradually to 0.76 GPa, a weak diamagnetic signal can be observed below ~1.7 K. At 1.02 GPa, a sharp superconducting transition in $\chi'(T)$ can be observed with about 60% superconducting volume fraction achieved at 1.5 K. With further increasing pressure, we can see a continuous increase of $T_c^{\chi'}$ from 2.86 K at 1.09 GPa to 4.29 K at 1.5 GPa, and the superconducting transition becomes much sharper at 1.5 GPa. Above 1.5 GPa, the superconducting transition was suppressed gradually, and the superconducting volume fraction reaches about 92% when the CDW order nearly vanishes.

Based on the above resistivity and magnetic susceptibility measurements under high pressures, we construct the $T$-$P$ phase diagram of RbV$_3$Sb$_5$ as shown in Figs. 4(a) and 4(b). From the phase diagram, we can easily visualize the evolution and intimated correlations between the CDW and SC as a function of pressure. With increasing pressure gradually, the CDW order is monotonically suppressed, accompanied by the initial enhancement of $T_c$ with a broad superconducting transition width $\Delta T_c \sim 2$ K below 1.5 GPa as displayed in Fig. 5(b) and 5(c), showing a strong competition between CDW and SC. At $P_{c1} \approx 1.5$ GPa, the highest $T_c^{zero} \approx 4$ K is achieved and it is over four times higher than that at ambient pressure. Above 1.5 GPa, the resistivity anomaly associated with CDW order becomes much weakened, while the superconducting transition temperature shows a shallow valley between 1.5 and 2.4 GPa. It seems that the long-range CDW order has been replaced by short-ranged one that coexists with SC in this pressure range and thus leads to a broadening of the superconducting transition, Fig. 5(b, c). The complete suppression of the short-ranged CDW order gives the second maximum of $T_c^{zero}$ around $P_{c2} \approx 2.4$ GPa. Above 2.4 GPa, the superconducting transition monotonically moves to lower temperatures and becomes very sharp with $\Delta T_c \approx 0.1$ K, Fig. 5(c).

To further probe the evolution of the superconducting electronic states of RbV$_3$Sb$_5$ under high pressures, we measured the upper critical field $\mu_0 H_{c2}$ at various pressures up to 5.2 GPa. All the $\rho(T)$ data under various magnetic fields and different pressures in PCC and CAC are shown in Fig. S2. $T_c$ moves to lower temperatures gradually with increasing magnetic fields. In order to track the evolution of $\mu_0 H_{c2}$, we employed the criteria of middle-point temperature $T_c^{mid}$ as the superconducting transition temperature. As shown in Fig. 5, we plot all $\mu_0 H_{c2}(T)$ data measured in PCC and CAC, and then estimate the zero-temperature $\mu_0 H_{c2}(0)$ by employing the empirical G-L equation as discussed above to fit the $\mu_0 H_{c2}(T)$ data. The best fitting results are indicated by the broken lines in Fig. 5(a, b). Surprisingly, the $\mu_0 H_{c2}(0)$ as a function of pressure exhibits a pronounced peak around $P_{c1} \approx 1.5$ GPa (Fig. 4(d)), but not at $P_{c2}$. This result is different from the double peak feature observed in CsV$_3$Sb$_5$ [34,39]. We also extract the initial slope of $\mu_0 H_{c2}(T)$ at each pressure, i.e., -d$H_{c2}$/d$T|_{Tc}$, and a similar peak feature shows around $P_{c1}$ (Fig. S3). As the slope -d$H_{c2}$/d$T|_{Tc}$ is proportional to the effective mass of charge carriers [47], and the divergence of -d$H_{c2}$/d$T|_{Tc}$ around $P_{c1} \approx 1.5$ GPa indicates an enhancement of effective mass as shown in Fig. 4(d). In general, the divergence of effective mass is considered as a hallmark of quantum criticality due to a complete suppression of certain electronic order [9,48]. It should be noted that the optimal superconducting phase usually emerges at the QCP in many unconventional superconductors [49-53].

*Discussions*.

By combining resistivity and magnetic susceptibility measurements, we have tracked the evolutions of the CDW order and SC in RbV$_3$Sb$_5$, and unveiled a shallow M-shaped double superconducting dome under pressure as described above. With increasing pressure, $T^*(P)$ decreases monotonically and vanishes completely around $P_{c2} \approx 2.4$ GPa, while $T_c(P)$ exhibits two maxima at $P_{c1} \approx 1.5$ GPa and $P_{c2}$, respectively. Above $P_{c2}$, the CDW order is completely eliminated and the superconducting transition shows a monotonic reduction. The highest $T_c^{onset} \approx 4.4$ K is achieved around $P_{c1} \approx 1.5$ GPa, rather than the putative QCP of the CDW order located at $P_{c2} \approx 2.4$ GPa. The optimal $T_c^{onset} \approx 4.4$ K around $P_{c1}$ is about fourfold enhanced in comparison with that at ambient pressure. All these characteristics in the T-P phase diagram of RbV$_3$Sb$_5$ are similar with those of the sister compound CsV$_3$Sb$_5$ [34], but having some quantitative differences between them. In addition, their double superconducting domes are also distinct from the observed single dome in KV$_3$Sb$_5$ under pressure [36].Thus, side-by-side comparisons among them are merited in order to have a better understanding on the unique properties of the $A$V$_3$Sb$_5$ family.

First of all, the character of the superconducting dome seems to correlate intimately with the $A$-cation size or the interlayer distance; the double-dome feature is weakened and changed to a single dome upon reducing the $A$-cation size from Cs though Rb to K. For CsV$_3$Sb$_5$, the larger Cs ion or interlayer distance should reduce the interlayer hopping and make the bands less dispersive along the $c$-axis. In principle, the highly two-dimensional character favors the formation of CDW order through the nesting scattering between van Hove points. Under high pressures, the bands become more dispersive along the $c$-axis with reducing the interlayer distance, and thus weaken the nesting scattering effect. The modification or vanishing of this out-of-plane CDW wavevector along the $c$ axis under pressure would give rise to the first SC dome around $P_{c1}$. In comparison with CsV$_3$Sb$_5$, the interlayer distance has been compressed chemically in RbV$_3$Sb$_5$, and thus the modification of CDW component along $c$-axis is expected to be weakened that would lead to a shallow M-shaped superconducting phase. While for KV$_3$Sb$_5$ with much smaller interlayer distance, the bands along the $c$-axis become more dispersive and therefore the double-dome character becomes much more weakened or even vanished as observed.

Secondly, although $T^*(P)$ displays monotonic suppression under physical pressure for these three compounds, the evolution of $T^*$ does not exhibit a similar trend as a function of $A$-cation size; it is peaked out at RbV$_3$Sb$_5$ with $T^* = 104$ K in comparison with that of 94 K for CsV$_3$Sb$_5$ and 78 K for KV$_3$Sb$_5$, respectively [12-14]. Accordingly, the superconducting $T_c$ at ambient pressure exhibits exactly opposite trend, illustrating a competition nature between these two intertwined orders. These comparisons highlight that the physical and chemical pressures should play some distinct roles in modifying the crystal and electronic structures that may need further investigations. Nonetheless, the critical pressures for the suppression of CDW order has a positive correlation with $T^*$ at ambient pressure; i.e. the corresponding critical pressures are decreased gradually from $P_{c1} \approx 1.5$ GPa and $P_{c2} \approx 2.4$ GPa for RbV$_3$Sb$_5$, to $P_{c1} \approx 0.6 - 0.9$ GPa and $P_{c2} \approx 2$

GPa for CsV$_3$Sb$_5$ [34], and finally to $P_{c1} \approx 0.4 – 0.5$ GPa for KV$_3$Sb$_5$ [36]. In addition, the optimal $T_c$ achieved under pressure also follows the same trend of $T_c$ at ambient.

Thirdly, the most interesting difference between CsV$_3$Sb$_5$ and RbV$_3$Sb$_5$ under pressure is the distinct behaviors of $\mu_0 H_{c2}(P)$ and its connection with the optimal $T_c$. For the former, the $\mu_0 H_{c2}(0)$ shows two peaks at $P_{c1}$ and $P_{c2}$ and the maximum $T_c$ emerges at $P_{c2}$ accompanying the complete suppression of CDW order; however, for the latter, both the $\mu_0 H_{c2}(0)$ and $T_c$ are peaked out at $P_{c1}$ rather than $P_{c2}$. Then, the question naturally arises why the maximal $\mu_0 H_{c2}(0)$ and $T_c$ in RbV$_3$Sb$_5$ do not show up at $P_{c2}$, which is expected to possess the strongest CDW fluctuations. Although more experiments are needed to clarify this issue, some hints from the experiments are noteworthy. That is, the observed shallower valley of the double superconducting dome in RbV$_3$Sb$_5$ indicates that the competition between CDW and SC in intermediate pressure range 1.5-2.4 GPa is much weakened in comparison with CsV$_3$Sb$_5$. As a result, the CDW fluctuations around $P_{c2}$ do not contribute significantly to the enhancement of $T_c$ as well as the electronic correlations.

Finally, it is noteworthy that the unusual double-dome superconducting phase observed in CsV$_3$Sb$_5$ and RbV$_3$Sb$_5$ is reminiscent of the phase diagrams of high-temperature cuprates [51,54,55] and FeSe-based superconductors [52], showing the presence of competing intertwined CDW/SDW or nematic orders. As indicated from the theoretical calculations [28-30], multiple electronic orders can be achieved as a function of on-site repulsion $U$ and nearest-neighbor Coulomb interaction $V$, such as ferromagnetism, intra-unit-cell antiferromagnetism, charge bond order or spin bond order. Thus, more experiments such as high-pressure nuclear magnetic resonance should be performed to further investigate the evolution of microscopic electronic orders in these V-based kagome metals.

## Conclusion

In summary, we have performed a comprehensive high-pressure study on RbV$_3$Sb$_5$ single crystals by employing the electrical transport and magnetic susceptibility measurements. At ambient pressure, the kagome metal RbV$_3$Sb$_5$ shows a charge order or CDW-like order at $T^* = 103$ K and SC at $T_c^{zero} = 0.78$ K. Our results reveal a subtle modification of the CDW order around $P_c \approx 1.5$ GPa, and the modified CDW is completely suppressed around 2.4 GPa. Correspondingly, the superconducting $T_c(P)$ displays the unusual M-shaped double superconducting dome structure with the optimal $T_c^{onset} \approx 4.4$ K and $T_c^{zero} \approx 4$ K at 1.5 GPa, and another maxima $T_c^{onset} \approx 3.93$ K and $T_c^{zero}$ ~ 3.8 K occurring at 2.4 GPa. Therefore, our phase diagram reveals the intimate interplay and strong competition between the CDW and SC in the pressure range 0 GPa $\leq P \leq 1.5$ GPa as evidenced by the broad superconducting transition width. Between 1.5 GPa and 2.4 GPa, the superconducting phase shows a valley character with possible underlying modification of the CDW. In addition, the $\mu_0 H_{c2}(0)$ shows a prominent peak character around 1.5 GPa, showing the characteristics of quantum criticality associated with the suppression of CDW order. The constructed $T$-$P$ phase diagram is similar to those of many unconventional superconductors with intertwined electronic orders and quantum criticality. Therefore, RbV$_3$Sb$_5$ together with CsV$_3$Sb$_5$ provide a new platform to study the correlations between the electronic instabilities and SC in this novel topological kagome metal family. In addition, the optimal $T_c$ of RbV$_3$Sb$_5$ reaches about 4.4 K at 1.5 GPa, which gives the possibility to further enhance $T_c$ of these V-based kagome superconductors. Further studies on RbV$_3$Sb$_5$ are need to address the open issues such as the character of CDW-like order in the intermediate pressure range.

## Acknowledgments

This work is supported by the National Natural Science Foundation of China (12025408, 11904391, 11921004, 11888101, 11834016, 11822412 and 11774423), the Beijing Natural Science Foundation (Z190008 and Z200005), the National Key R&D Program of China (2018YFA0305700, 2018YFE0202600 and 2016YFA0300504), the Strategic Priority Research Program and Key Research Program of Frontier Sciences of the Chinese Academy of Sciences (XDB25000000, XDB33000000 and QYZDB-SSW-SLH013), and the CAS Interdisciplinary Innovation Team. U.W. is supported by the JSPS KAKENHI Grant No. 19H00648.

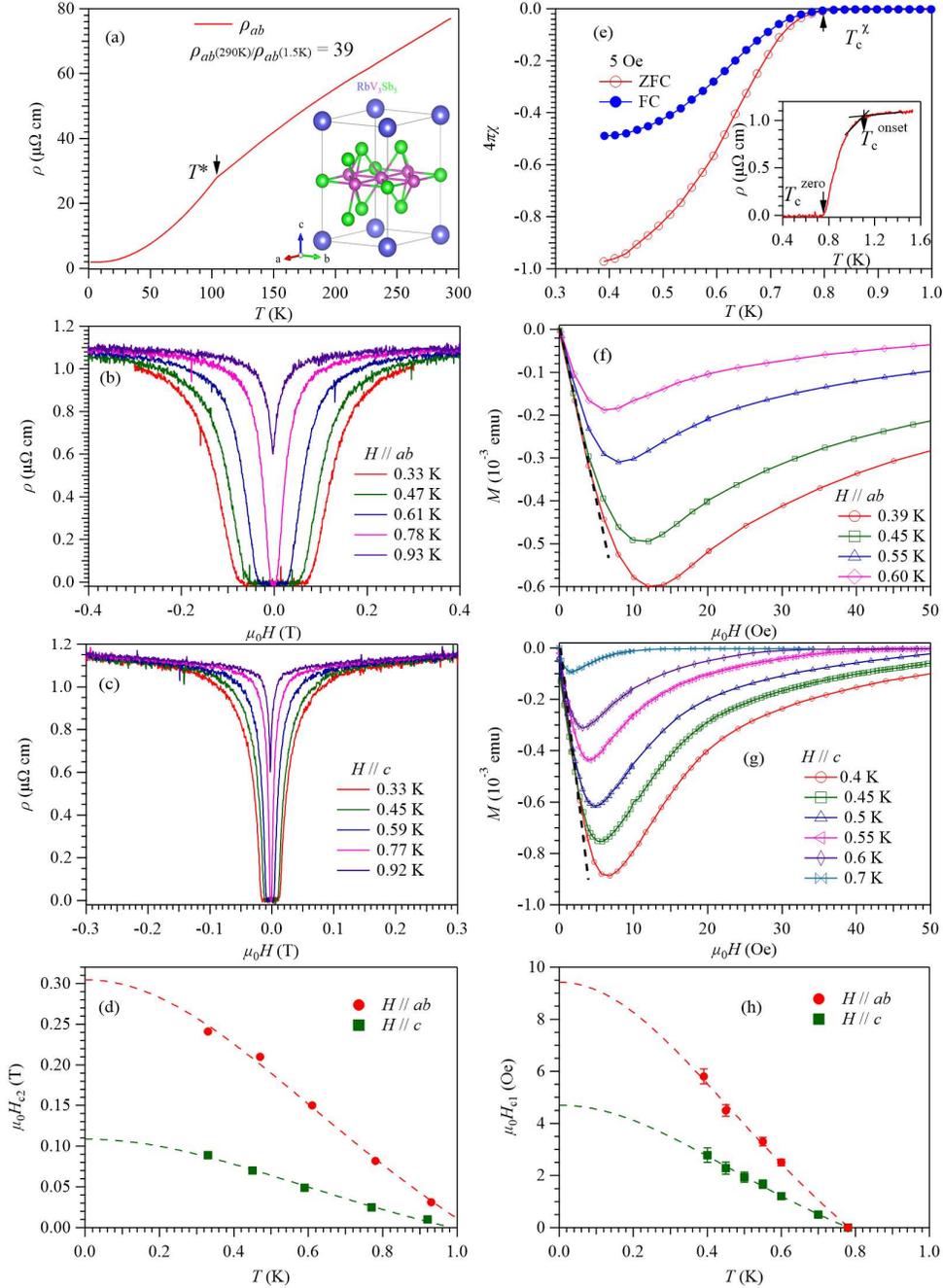

Figure 1: (a) In-plane resistivity $\rho(T)$ of RbV$_3$Sb$_5$ at ambient pressure from 300 K down to 2 K. (b)-(c) Field dependences of $\rho(T)$ measured with field parallel to *ab* plane and *c*-axis at various temperatures. (d) The anisotropic upper critical field $\mu_0H_{c2}$, defined as the field at 90% of the normal-state resistivity. (e) The dc magnetic susceptibilities under zero-field-cooled (ZFC) and field-cooled (FC) conditions. The inset shows the $\rho(T)$ below 1.6 K highlighting the superconducting transition. (f)-(g) Isothermal magnetization at various temperatures with magnetic field parallel to *ab* plane and *c*-axis. (h) The anisotropic lower critical field $\mu_0H_{c1}$, defined as the fields at which *M-H* curves start to deviate from the linear line indicated by the dashed lines in (f) and (g).

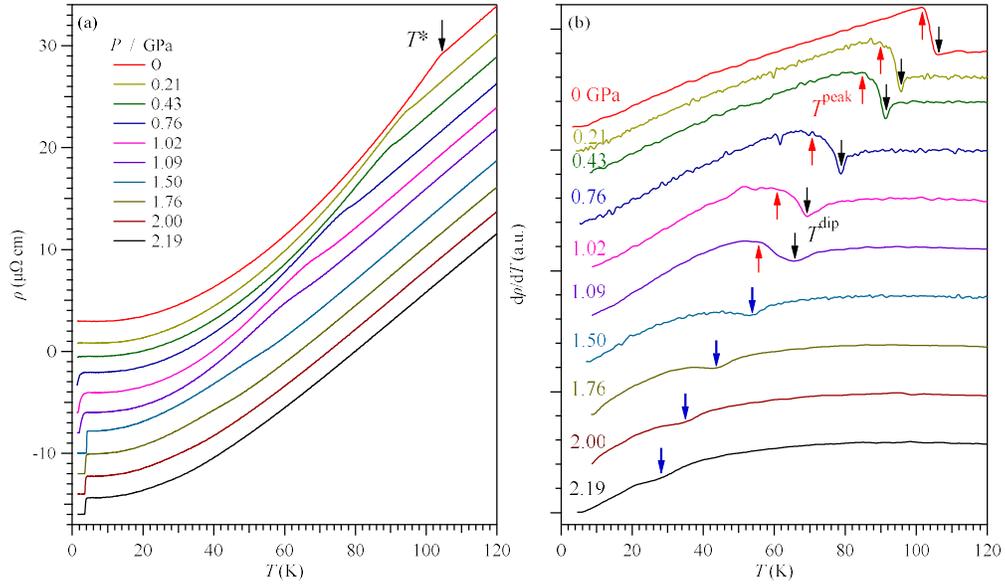

Figure 2. Variation of the charge-order-related transition under high pressures. Temperature dependences of (a) resistivity $\rho(T)$ and (b) its derivative $d\rho/dT$ for the RbV$_3$Sb$_5$ sample measured in piston cylinder cell (PCC) under various pressures up to 2.19 GPa. The charge-order or CDW-like transition temperature $T^*$ are marked by the arrows in the figures. The curves in (a) and (b) have been shifted vertically for clarity.

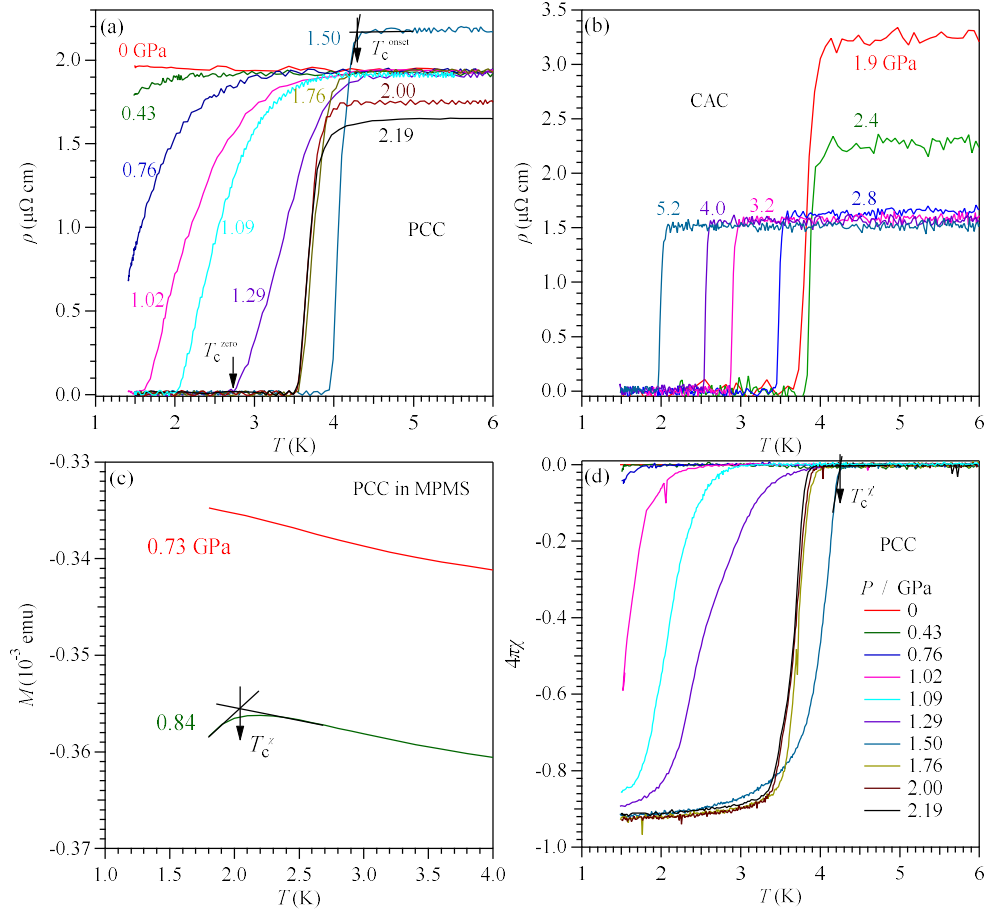

Figure 3. Variation of the superconducting transition under high pressures in (a, b) resistivity and (c, d) magnetic susceptibility. The resistivity $\rho(T)$ data in (a) and (b) are measured up to 2.19 GPa with a piston-cylinder cell (PCC) and up to 5.2 GPa with a cubic anvil cell (CAC), respectively. The dc magnetization in (c) was recorded in MPMS with a miniature PCC, while the ac magnetic susceptibility in (d) was measured with the mutual induction method in PCC.

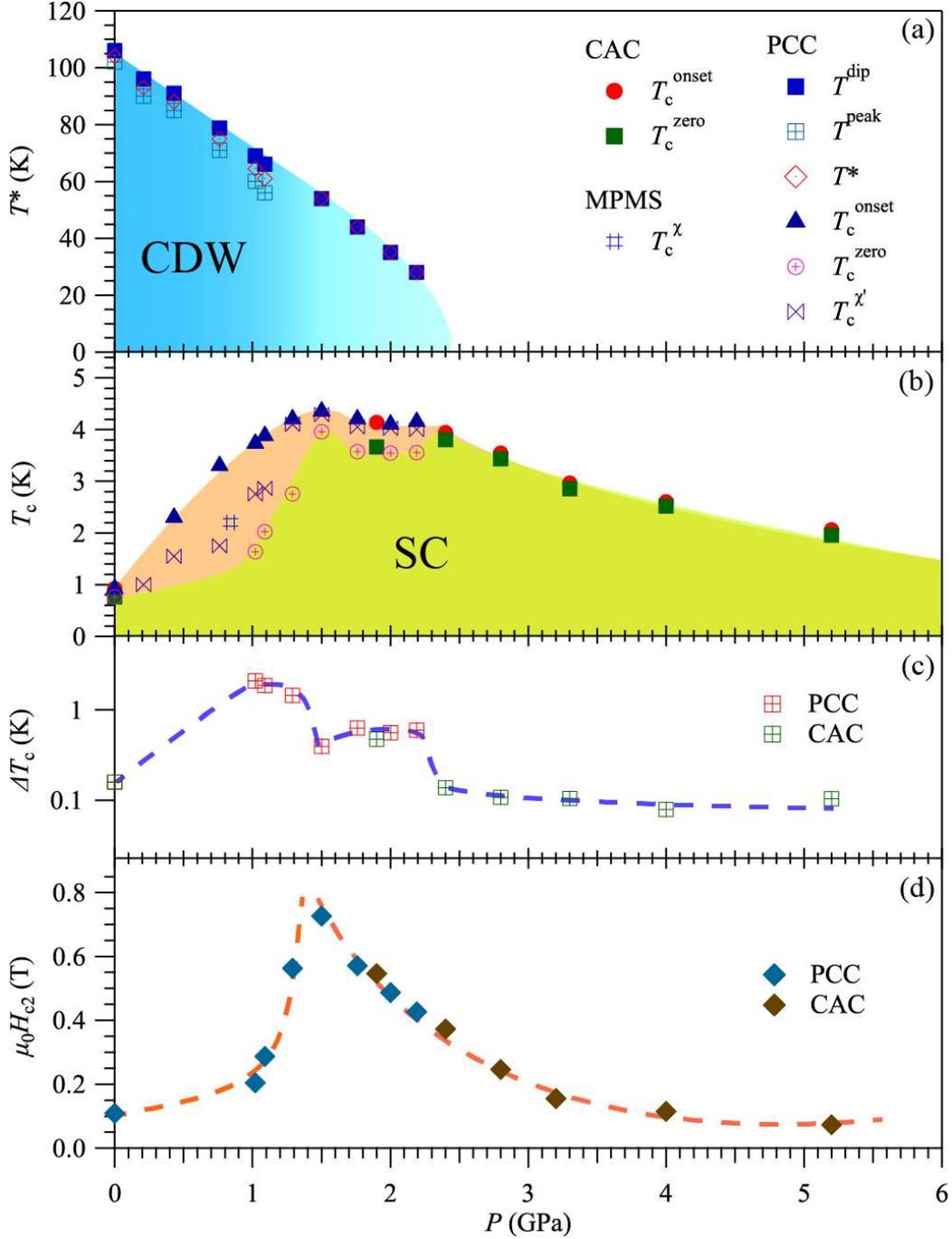

Figure 4. Temperature-pressure phase diagram of $CsV_3Sb_5$. Pressure dependences of (a) $T^*$, (b) $T_c$, and (c) the superconducting transition width $\Delta T_c$ determined from the resistivity and magnetic measurements on several samples, and (d) the zero-temperature upper critical field $\mu_0 H_{c2}(0)$ obtained from the empirical Ginzburg–Landau (GL) fitting.

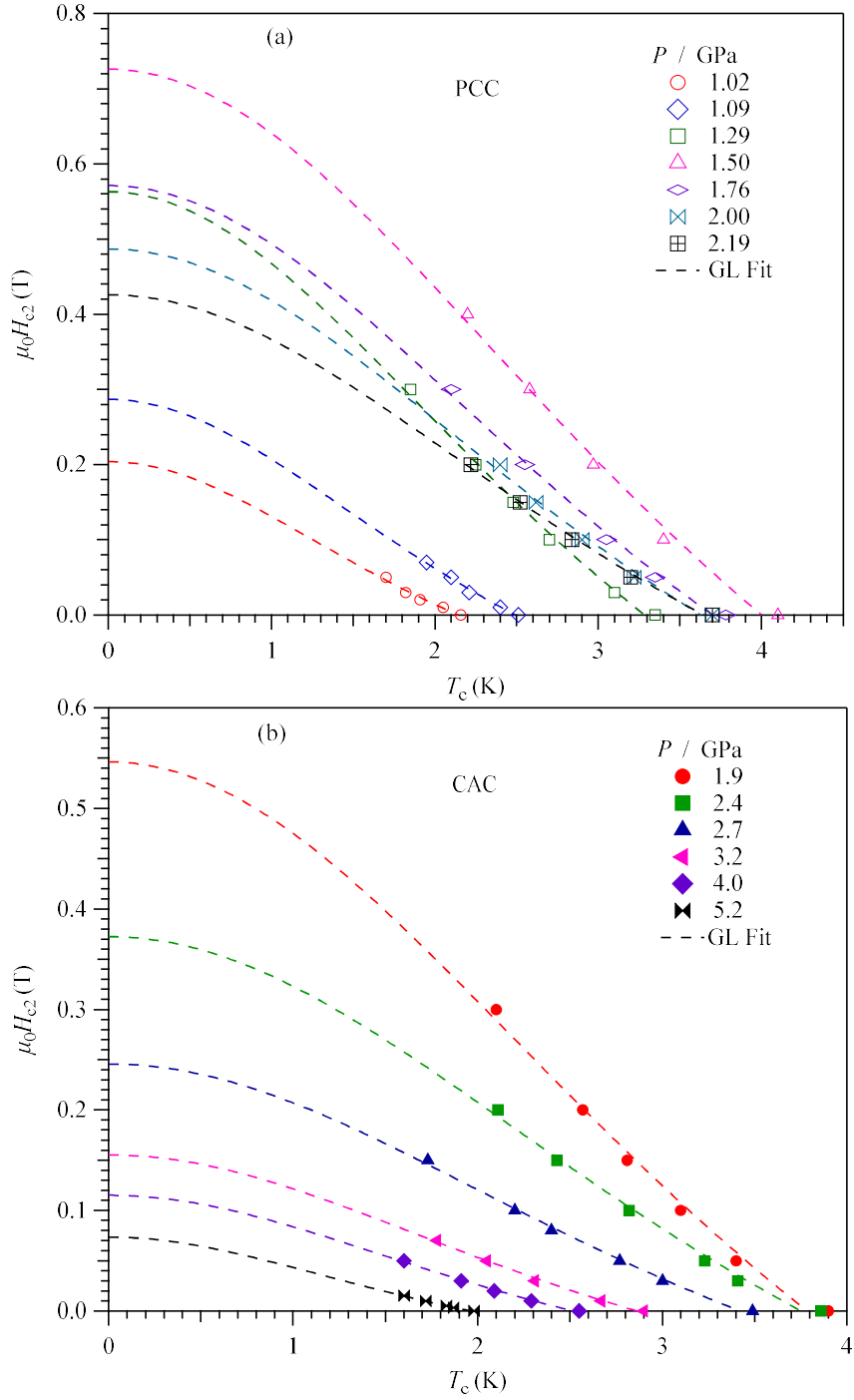

Figure 5. Temperature dependences of the upper critical field $\mu_0 H_{c2}$ at different pressures measured with (a) PCC and (b) CAC. The broken lines represent the Ginzburg–Landau (GL) fitting curves.